 \documentclass[sigconf]{acmart}

\AtBeginDocument{%
  \providecommand\BibTeX{{%
    \normalfont B\kern-0.5em{\scshape i\kern-0.25em b}\kern-0.8em\TeX}}}

\copyrightyear{2025}
\acmYear{2025}
\setcopyright{rightsretained}
\acmConference[CHIWORK '25]{CHIWORK '25: Proceedings of the 4th Annual Symposium on Human-Computer Interaction for Work}{June 23--25, 2025}{Amsterdam, Netherlands}
\acmBooktitle{CHIWORK '25: Proceedings of the 4th Annual Symposium on Human-Computer Interaction for Work (CHIWORK '25), June 23--25, 2025, Amsterdam, Netherlands}
\acmPrice{}
\acmDOI{10.1145/3729176.3729179}
\acmISBN{979-8-4007-1384-2/25/06}

\usepackage{caption,subcaption}
\usepackage[titletoc,title]{appendix}
\usepackage{siunitx}
\usepackage{xcolor}
\usepackage{soul}

\usepackage{diagbox}
\usepackage{booktabs}

\newcommand{\za}[1]{\textcolor{black}{#1}}

\begin{document}

\title{Emerging Reliance Behaviors in Human-AI Content Grounded Data Generation: The Role of Cognitive Forcing Functions and Hallucinations}

\author{Zahra Ashktorab}
\email{zahra.ashktorab@ibm.com}
\affiliation{%
  \institution{IBM Research}
  \city{Yorktown Heights}
  \state{NY}
  \country{USA}
}
\author{Michael Desmond}
\email{michael.desmond@ibm.com}
\affiliation{%
  \institution{IBM Research}
  \city{Yorktown Heights}
  \state{NY}
  \country{USA}
}
\author{Qian Pan}
\email{qian.pan@ibm.com}
\affiliation{%
  \institution{IBM Research}
  \city{Cambridge}
  \state{MA}
  \country{USA}
}

\author{James Johnson}
\email{jmjohnson@us.ibm.com}
\affiliation{%
  \institution{IBM Research}
  \city{Cambridge}
  \state{MA}
  \country{USA}
}
\author{Michelle Brachman}
\email{michelle.brachman@ibm.com}
\affiliation{%
  \institution{IBM Research}
  \city{Cambridge}
  \state{MA}
  \country{USA}
}

\author{Casey Dugan}
\email{cdugan@rand.org}
\affiliation{%
  \institution{Rand Corporation}
  \city{Cambridge}
  \state{MA}
  \country{USA}
}

\author{Marina Danilevsky}
\email{mdanile@us.ibm.com}
\affiliation{%
  \institution{IBM Research}
  \city{Almaden}
  \state{CA}
  \country{USA}
}
\author{Werner Geyer}
\email{werner.geyer@ibm.com}
\affiliation{%
  \institution{IBM Research}
  \city{Cambridge}
  \state{MA}
  \country{USA}
}

\renewcommand{\shortauthors}{Ashktorab et al.}

\renewcommand{\shorttitle}{Emerging Reliance Behaviors in Human-AI Content Grounded Data Generation}

\begin{abstract}
We investigate the impact of hallucinations and Cognitive Forcing Functions in human-AI collaborative content grounded data generation, focusing on the use of Large Language Models (LLMs) to assist in generating high-quality conversational data. Through a study with 34 users who each completed 8 tasks (n=272), we found that hallucinations significantly reduce data quality. While Cognitive Forcing Functions do not always alleviate these effects, their presence influences how users integrate AI responses. Specifically, we observed emerging reliance behaviors, with users often appending AI-generated responses to their correct answers, even when the AI’s suggestions conflicted. This points to a potential drawback of Cognitive Forcing Functions, particularly when AI suggestions are inaccurate. Users who overrelied on AI-generated text produced lower-quality data, emphasizing the nuanced dynamics of overreliance in human-LLM collaboration compared to traditional human-AI decision-making.
\end{abstract}

\keywords{Large Language Models, Hallucinations, Human-AI Interaction, Cognitive Forcing Functions, Data Quality, Overreliance}

\ccsdesc[500]{Human-centered computing~Empirical studies in HCI}

\maketitle

\section{Introduction}

Recent advances in AI technology have led to widespread adoption of LLMs across various contexts, from customer support to personal assistants. LLMs are capable of generating text at a remarkable level of coherence and fluency. For use of LLMs in more specialized domains \cite{gururangan2020don} and language tasks \cite{kenton2019bert}, fine-tuning pre-trained language models has been the dominant methodology.  Fine tuning involves updating the model's weights and parameters based on the new data while retaining the knowledge learned during its initial training \cite{fan2023recommender}. However, creating this data can be a challenge. 
 
 For content-grounded dialog tasks, a pair of annotators are required to converse over a topic contained within a grounding document. This is a slow and expensive process. A more scalable approach is to employ a single annotator who works directly with an LLM to create the dialog. The annotator plays the role of the user asking questions, and the model generates candidate responses based on the grounding document. The annotator, who also has access to the document, then corrects and edits the model responses when necessary to produce high quality training data. This process represents a shift in the nature of annotation work—rather than merely labeling or transcribing data, annotators now act as co-creators, refining and shaping AI-generated content. To support this new type of work, guidelines are provided to help annotators systematically correct and enhance model outputs, typically by improving accuracy, coherence, and informativeness. This shift is particularly relevant in domains that require specialized knowledge for fine-tuning, such as human resources (HR), information technology (IT), customer support, healthcare, and legal services, where domain expertise is crucial for ensuring high-quality AI-generated interactions. As LLMs continue to be integrated into training data pipelines, this evolving role of annotators raises important questions about expertise, labor dynamics, and human-AI collaboration in data generation.

 
 Using AI in dialog creation can introduce the well-researched problem of automation bias in Human-AI systems. Unlike errors in other AI systems, LLMs can not only produce inaccurate information but may even hallucinate new information \cite{maynez2020faithfulness}, which annotators may inadvertently accept a model's response when it is incorrect, thus overrelying on the AI generated responses. Users may not edit the responses adequately, leading to low-quality data that may not accurately reflect real-world conversational interactions. To improve the quality of collected fine-tuning data in the context of content grounded dialog, we test a set of features that aim to encourage users to think critically about generated model responses, reducing the risk of overreliance on AI. We rely on prior work in the AI-decision making that explored the role of Cognitive Forcing Functions, interventions applied at the decision making point that disrupt routine thinking prompting analytical reasoning \cite{lambe2016dual} on overreliance \cite{buccinca2021trust}. In this paper, we conduct a study focusing on the impact of Cognitive Forcing Functions and hallucinations in turn-by-turn grounded conversational contexts, where data creation plays a critical role in fine-tuning conversational AI models. \za{Specifically, we focus on the use of LLMs to assist data workers to generate high quality conversational data for fine tuning.} We investigate the influence of hallucinated text generated by AI on data quality.  In the context of data for fine-tuning conversational AI models, our research questions are as follows: 
\label{sec:RQs}
\begin{description}
\item \textbf{RQ1} \za{How do hallucinations and Cognitive Forcing Functions affect the quality of text produced in content grounded data generation?}

\item \textbf{RQ2} \za{How do hallucinations and Cognitive Forcing Functions affect user reliance in content grounded data generation?}

\item \textbf{RQ3} \za{What types of reliance behaviors are observed in content grounded data generation and how do these behaviors impact data quality? }
\end{description}

 By addressing these questions, we hope to provide insights into how to improve the quality of conversational data and promote more natural and informative interactions between users and AI agents. Our paper makes the following contributions. 
 \begin{itemize}
 \item We present a rubric for evaluating content grounded dialogue for customer support. This rubric spans faithfulness, accuracy, completeness, and is applicable across content-grounded dialog contexts. 
 \item We designed and present three Cognitive Forcing Functions in the context of this content-grounded dialog creation task: Highlighting, Formulate, Read First.
 \item Cognitive Forcing Functions (CFFs) influenced AI reliance and data quality, even without actively mitigating hallucination effects. While CFF Presence did not significantly reduce the negative impact of hallucinations, it improved overall data quality and shaped user behavior. Notably, in the Formulate condition, participants incorporated AI-generated responses more when the CFF was absent, likely as a cognitive offloading strategy to compensate for prior effort. 
 \item We identify emerging reliance behaviors afforded by the text-generation context. Specifically, we observed that users append AI responses to their already correct responses, even when the information is conflicting, demonstrating a potential downside to use of Cognitive Forcing Functions, especially when the AI suggestion is incorrect.
 \end{itemize}

\section{Related Work} 
\subsection{Human AI Collaboration \za{ and Decision Making}}
\za{Previous research in Human-Computer Interaction (HCI) has explored how different factors influence the outcomes of human-AI interactions. One recent study explored the cognitive processes in human-AI decision making when dealing with uncertainty. Participants were exposed to varying levels of uncertainty when interacting with various decision support systems. The findings showed that the presence of uncertainty information can lead to productive skepticism and ultimately reduce users' overreliance on decision support systems \cite{prabhudesai2023understanding}. Another study highlighted data scientists' use of interpretability tools in machine learning and the potential for misuse and over-trust in AI, particularly when users don't fully understand the tools' capabilities or limitations. The study found that there was often a mismatch between users mental models of the tool's abilities and the tool's actual capabilities \cite{kaur2020interpreting}.  AI tools can greatly aid decision-making, especially for less experienced users, but the effect varies based on the user's prior knowledge and the AI's tuning. A study exploring the dynamics of human-AI collaboration in a non-trivial blood vessel labeling task where participants indicated whether a given blood vessel was flowing or stalled found AI recommendations can improve users accuracy, especially for those with lower baseline performance. High-performers maintained their accuracy regardless of AI presence. The study also explored agreement with the AI-assistant and found that the likelihood of agreement with AI recommendations varied and users were more likely to agree with incorrect recommendations in the high true positive rate condition when compared to the other two tuning conditions (high true negative rate and balanced condition).   \cite{inkpen2023advancing}. Perceptions of AI explanations are heavily influenced by one's background in AI, as demonstrated in the research on how AI backgrounds shape perceptions of AI explanations that demonstrated both those with AI and without AI backgrounds exhibited more faith in numerical explanations (compared to natural language with justification and natural language without justification) but for different reasons \cite{ehsan2021explainable}. These factors collectively underscore the complexity of human-AI interactions and the need for careful consideration of the design of AI systems to optimize decision-making processes.}

\subsubsection{\za{LLMs}}
LLMs have demonstrated remarkable capabilities of generating natural language \cite {brown2020language}. One limitation of large language models is there is still a need for task-specific fine-tuning data for model improvement to achieve strong performance for specific tasks. Fine-tuning data can consist of large amount of specific examples of that task \cite{brown2020language}. Fine-tuning an LLM involves the process of augmenting an existing language model with additional data specific to the subject matter at hand, grounding the model in that data. The weights of a pre-trained model are updated by training on a dataset that is task specific \cite{fan2023recommender}. This augmentation enables the LLM to specialize in a particular domain or improve its performance on a specific task. The fine-tuning process often relies on large datasets annotated by human annotators, providing curated input to train the LLM effectively.  The text generated from large language models have become so advanced that it is indistinguishable from human-generated text \cite{jakesch2023human}. Many heuristics people use to distinguish AI-generated from human-generated are flawed like belief that presence of grammatical issues or long words is associated by AI-generated, when it is more associated with human-generated content \cite{jakesch2023human}. In the context of human-LLM interaction, reliance on AI takes a more complex form and may not just mean simply agreeing with the AI text generation. The interaction comprises of many formats including: keeping the AI-generated text in its entirety, adding minor edits, removing content, or implicit influence by the AI-generated text. In many recent HCI studies, researchers have used various strategies to evaluate the resulting text generated from human-LLM collaboration. Jakesch et al. used a set of 500 annotators to evaluate whether text co-written by humans and an AI writing assistant argued for or against social media. Each crowdworker evaluated 25 responses and split the resulting text into segments to determine whether it argued in favor of social media use, against social media use or for both in favor and both against \cite{10.1145/3544548.3581196}. Other work has also shown that through human evaluation, the text generated from language models have become so advanced that it is indistinguishable from human-generated text \cite{jakesch2023human}.

\subsubsection{Faithfulness Evaluation}
One important aspect to consider when examining AI-generated responses is the phenomenon of hallucinations. Hallucinations are unique to Language Models (LLMs) and refer to non-factual statements generated by AI. This sets them apart from ML predictions, such as sentiment analysis or decision-making, which are based on large and often broad datasets. AI-generated summaries have been known to contain information not in context \cite{maynez2020faithfulness}. A judge even sanctioned a lawyer who used a hallucinated case by ChatGPT to support his legal argument only to find that the case was hallucinated \cite{tenzer2023defamation}. Hallucinations generated by AI can be especially difficult to detect when they appear to be  plausible and relevant to the given context but are factually incorrect.
Faithfulness, in the context of AI-generated responses, refers to the degree to which the generated output aligns with the content and information present in the reference document. Faithfulness is described as a key indicator for the quality of an LLM-generated summarization \cite{maynez2020faithfulness}. Hallucinations, as described above, demonstrate low faithfulness because the information presented in the responses is not grounded in the reference document. Evaluation of faithfulness has long been a challenge in the text summarization domain. \cite{wang2022analyzing}.  Automatic methods like ROUGE \cite{lin2004rouge} that check for faithfulness of the summary to the original text often do not match human judgements \cite{fischer2022measuring}. Another method of hallucination detection is to leverage existing intrinsic uncertainty metrics like entropy \cite{yuan2021bartscore}. Some of these methods may require access to token-level probability distribution that is not available to users who are accessing LLMs through APIs. Prior work on fact-verification requires an external database of knowledge to test the factuality of a claim \cite{thorne2018fact}. Beyond hallucinations, recent work highlights that AI-generated misinformation can differ significantly from human-written misinformation in style and presentation, often sounding more detailed, uncertain, or personal—yet still perceived as credible—posing challenges for existing evaluation methods and detection systems \cite{10.1145/3544548.3581318}. Distinguishing between human- and AI-generated content across creative and technical genres like stories, poetry, and code remains a growing challenge, with models performing better on simpler classification tasks but struggling to identify more sophisticated outputs from advanced LLMs such as GPT \cite{hayawi2024imitation}.Faithfulness evaluation has long been a consideration for text generated by content grounded LLMs.


%

\subsection{Reliance on AI}
The adoption of AI systems in various decision-making domains has become increasingly common. Despite the remarkable accuracy of AI, human-AI teams sometimes fall short of expectations \cite{campero2022test}, particularly in areas where AI-alone teams outperform human-alone counterparts across various domains  \cite{jacobs2021machine,green2019principles,10.1145/3377325.3377498,zhang2020effect,buccinca2021trust}. One plausible explanation for these performance disparities is the phenomenon of human overtrust in AI. When users excessively trust the system, they may accept the AI's suggestions even when they are incorrect. Extensive work has been done on achieving calibrated trust in AI systems \cite{zhang2020effect, tomsett2020rapid}. Calibrated trust in AI signifies that users have an appropriate level of trust in a system that accurately reflects its capabilities and performance. Many researchers have explored whether enhancing transparency can improve calibrated trust in AI. However, there is limited evidence to suggest that transparency improves trust, whether through the use of interpretable models or by allowing users to inspect model behavior, providing explanations, or reducing the number of features presented \cite{poursabzi2021manipulating, kunkel2019let, 10.1145/3290605.3300789}.

Automation bias, also known as overreliance, is defined as the ''tendency to over-rely'' on automated systems \cite{goddard2012automation}. It occurs when individuals unquestionably follow recommendations from automated systems without applying their own perceptual abilities and judgment. Overreliance has been extensively studied in critical industries, including medicine \cite{goddard2012automation} and aviation \cite{skitka2000automation}.Researchers have examined reliance on artificial intelligence  from various angles. Reliance has been defined as as accuracy of a decision based on correct or incorrect AI advice \cite{10.1145/3411764.3445717, 10.1145/3397481.3450650}. It has also been defined as a ratio, either ratio of following versus not following incorrect AI advice \cite{buccinca2021trust}, or ratio of following correct or incorrect AI advice \cite{jakubik2022empirical, 10.1145/3377325.3377480}. Schemmer et al. introduced the concept of AoR (Appropriateness of Reliance) as a quantifiable two-dimensional measurement that considers (correct/incorrect) self-reliance as well as (correct/incorrect) AI reliance \cite{schemmer2023appropriate}.  We carry over these definitions by redefining hallucinated responses as incorrect advice in our context and measuring whether users follow incorrect advice or do not follow correct advice.


\subsection{Cognitive Forcing Functions}
The dual processing theory suggests that human decision-making can be categorized into two distinct cognitive processes: fast and automatic (System 1 thinking) and slow and deliberate (System 2 thinking) \cite{kahneman2011thinking}.  Transitioning individuals from System 1 to System 2 thinking presents a challenge that researchers have tackled across various domains. Annotation tasks which are repetitive are prone to System 2 thinking \cite{ashktorab2021ai}. Bu\c{c}inca et al. examined overreliance in the context of AI decision-making systems. They developed three Cognitive Forcing Functions designed to shift users from System 1 thinking to System 2 thinking, encouraging more critical thought. These CFFs include asking individuals to make decisions before viewing the AI's recommendation and slowing down the decision-making process, allowing individuals to choose whether and when to access the AI recommendation \cite{buccinca2021trust}. These Cognitive Forcing Functions draw inspiration from different psychological theories. For instance, the CFF that prompts users to decide before seeing the AI's recommendation is rooted in the anchoring bias, which arises when individuals are influenced by the initial AI recommendation before forming their own judgment \cite{green2019principles}. Delaying the presentation of AI-generated suggestions is based on previous research suggesting that postponing the AI's recommendation can potentially enhance decision outcomes \cite{park2019slow}. Moreover, providing individuals with the choice to decide whether they want to engage with the Cognitive Forcing Function aligns with the concept of reactance or resistance to advice. Reactance occurs when unsolicited advice triggers an individual's initial ideas and resistance \cite{fitzsimons2004reactance}. Their findings revealed that Cognitive Forcing Functions were more effective in mitigating overreliance compared to AI-generated explanations or the baseline condition. However, users assigned the lowest subjective ratings to the Cognitive Forcing Functions that were most effective at reducing overreliance \cite{buccinca2021trust}. Thus, a question remains unanswered in the context of AI-generated text: How can Cognitive Forcing Functions be best designed, and are they effective in situations where users rely on AI-generated text for content creation? Our study distinguishes itself by focusing on AI-generated text for content creation rather than the decision making context, tailoring Cognitive Forcing Functions specifically for this context. Our research addresses a gap in AI-generated text for content creation, offering insights distinct from existing studies \cite{buccinca2021trust}.

\section{Background}
\label{sec:background}
Within a large, multinational company, Human Resources (HR) is looking to automate some of the question-answering that takes up much of the time of employees within that department. Using LLMs to do this is attractive given recent advances, but it is well known that they hallucinate and must be fine-tuned to avoid potential misinformation in HR-related tasks and associated liability concerns. Our study's design was shaped by our need to generate high quality fine-tuning data for content-driven dialogue tasks. This need extends beyond HR to various customer support domains lacking pre-existing content-grounded dialogue (i.e. a company launching a new product with associated documentation).  While the subject of our inquiry is HR, we use Customer Support and HR support interchangeably in the study. Customer support's primary aim is to assist customers by addressing inquiries and concerns about products or services, ensuring a positive experience and satisfaction. In a large corporation, HR services oversee workforce management, including payroll and benefits administration. Employees often approach HR with questions about various benefits and services, including healthcare similar to how customers might approach customer services about products or services \cite{majumder2021chatbots,vosburgh2007evolution,hayward2006transforming}. 


\subsection{Co-creating Content Grounded Data for AI-Assisted Customer Support}
\label{sec:create}
The training dataset required to fine-tune a LMM is task-specific. For example, in a Spanish to English translation task, the model is given a large corpus of examples of Spanish to English translations. Similarly, for a content-grounded dialog task, a model would be given grounding documents and associated conversations between a human and an agent specific to those documents. To curate this data its possible to employ a pair of human annotators, one who plays the role of the customer and another who plays the role of the customer support agent to have a hypothetical conversation. Below illustrates the task those two humans could be given: 

\begin{enumerate}
    \item Customer has recently joined company. Wants to know about what company covers regarding working from home. Come up with one or two items customer is specifically interested in, such as getting home internet expenses reimbursed, or buying an ergonomic chair.
\item Customer has children and wants to know if their place of employment offers any child care credit benefits. Have the children be different ages, for example: one in preschool (under 5), one in elementary school, one in high school / college (over 13). And/or, have the customer be a parent-to-be who will soon have a new baby (either through customer or partner giving birth, or adoption).
\end{enumerate}

During this data creation process, the human annotators may be given guidelines and a reference document that helps them craft appropriate questions and responses. The guidelines may emphasize the importance of maintaining politeness and conversational tone (by the agent) to ensure a positive user experience.  However, having two humans interact is slow and expensive. A more scalable solution is to have an LLM play the role of the customer support agent, prompting it to generate the next agent response given the reference document and the dialog history. From the annotators point of view, they would ask questions and then would have the ability to edit the LLM/agents response to ensure that it is accurate, faithful to the reference document and conversational in nature. Each turn of this dialog can then be used as a training example to fine-tune an LLM to perform customer support tasks. 


Developing fine-tuning data with the help of an LLM producing candidate responses reduces the amount of annotator effort, however the annotator may over-rely on the model generated responses which may adversely affect the quality of the training data. 
The potential risk of relying too heavily on AI responses arises from the fact that models are not infallible. AI models can and do produce incorrect data or data that is not grounded in the document. 

\subsection{Defining Hallucinations}
\label{sec:hallucinations}

In the customer support context, there are instances where the AI might generate a response that is highly relevant to the user's query but has no basis in the reference document.  We can illustrate this concept with an example scenario. Consider a reference document that provides information on  what a large corporation offers in terms of short term disability benefits. A customer poses the following question: \textit{"I am injured and cannot return to work. I just applied for short term disability. How long is short term disability?"} The AI system could generate several different responses, and we will label them as Model A, Model B, and Model C. 

\begin{itemize}
  \item \textbf{Model A:} \textit{"26 weeks within one year period. Your manager has to approve the application."}
  \item \textbf{Model B:} \textit{"You get 13 weeks as an exemption due service."}
  \item \textbf{Model C:} \textit{"You get 1-month renewable every 4 Months"}
\end{itemize}

At first glance, all of these responses may appear to be correct, as they provide relevant information about short term disability income. However, in this example, they are all hallucinations because the response is not aligned with the relevant content of the reference document.  These hallucinations can refer to language from the reference document that do not answer the customer query. These model responses were generated with non-fine-tuned models, using the reference document as the prompt. It would be unethical and dangerous if incorrect responses were given to employees on topics relating to health insurance, life insurance and other benefits. For the purposes of our study, we were particularly interested in exploring responses that exhibit high relevance to the user's query while showing low faithfulness to the document. Such examples are challenging for users to identify as they seem plausible and appropriate based on the query, but they lack support from the underlying document. Our Cognitive Forcing Functions attempt to push people into thinking more critically (System 2 thinking) before accepting potentially hallucinated responses.

\section{Hypotheses}
We hypothesize that hallucinations adversely affect text quality, diminishing faithfulness, accuracy, and completeness (H1) and decrease reliance on the AI (H3), or the degree to which the users leveraged the AI suggested response when formulating their own responses. These hypotheses are informed by studies showing the impact of incorrect AI outputs on decision-making processes \cite{alufaisan2021does}.  Similarly, Cognitive Forcing Functions have been effective in improving decision-making outcome in human-AI interaction \cite{buccinca2021trust}. These functions could potentially lead users to more critically evaluate AI-generated text leading to various quality outcomes. For example,  \cite{buccinca2021trust} found that forcing the user make a decision before seeing the AI suggested decision mitigates overreliance in decision making outcomes. Similarly, the Cognitive Forcing Function forces users to create a response before seeing the AI response in a text generation context. We hypothesize that results from \cite{buccinca2021trust} generalize from the decision making context to the text-generation context and so we will see high data quality from our Cognitive Forcing Function (H2) and less use of the AI response when a hallucination is present (H4).

\begin{description}
\item \textbf{H1} Hallucinations decrease data quality.
\item \textbf{H2} All Cognitive Forcing Functions, when present will mitigate the detrimental impact of hallucinations on data quality. 
\begin{description}
\item \textbf{H2A} The Formulate Cognitive Forcing Function, when present will mitigate the detrimental impact of hallucinations on data quality. 
\item \textbf{H2B} The Read First Cognitive Forcing Function, when present will mitigate the detrimental impact of hallucinations on data quality. 
\item \textbf{H2C} The Highlight Cognitive Forcing Function, when present will mitigate the detrimental impact of hallucinations on data quality. 
\end{description}

\item \textbf{H3} Hallucinations decrease user reliance on AI.
\item \textbf{H4} All Cognitive Forcing Functions, when present, will decrease use of AI when hallucinations are present compared to compared to when the Cognitive Forcing Functions are absent.
\begin{description}
 \item \textbf{H4A} The Formulate Cognitive Forcing Function, when present, will decrease use of AI when hallucinations are present compared to when the Cognitive Forcing Function is absent.
 \item \textbf{H4B} The Read First Cognitive Forcing Function, when present, will decrease use of AI when hallucinations are present compared to when the Cognitive Forcing Function is absent.
 \item \textbf{H4C} The Highlight Cognitive Forcing Function,
 when present, will decrease use of AI when hallucinations are present compared to when the Cognitive Forcing Function is absent. 
\end{description}
\end{description}

\section{Methodology}

\begin{figure*}[ht]
  \centering
\includegraphics[width=1\linewidth]{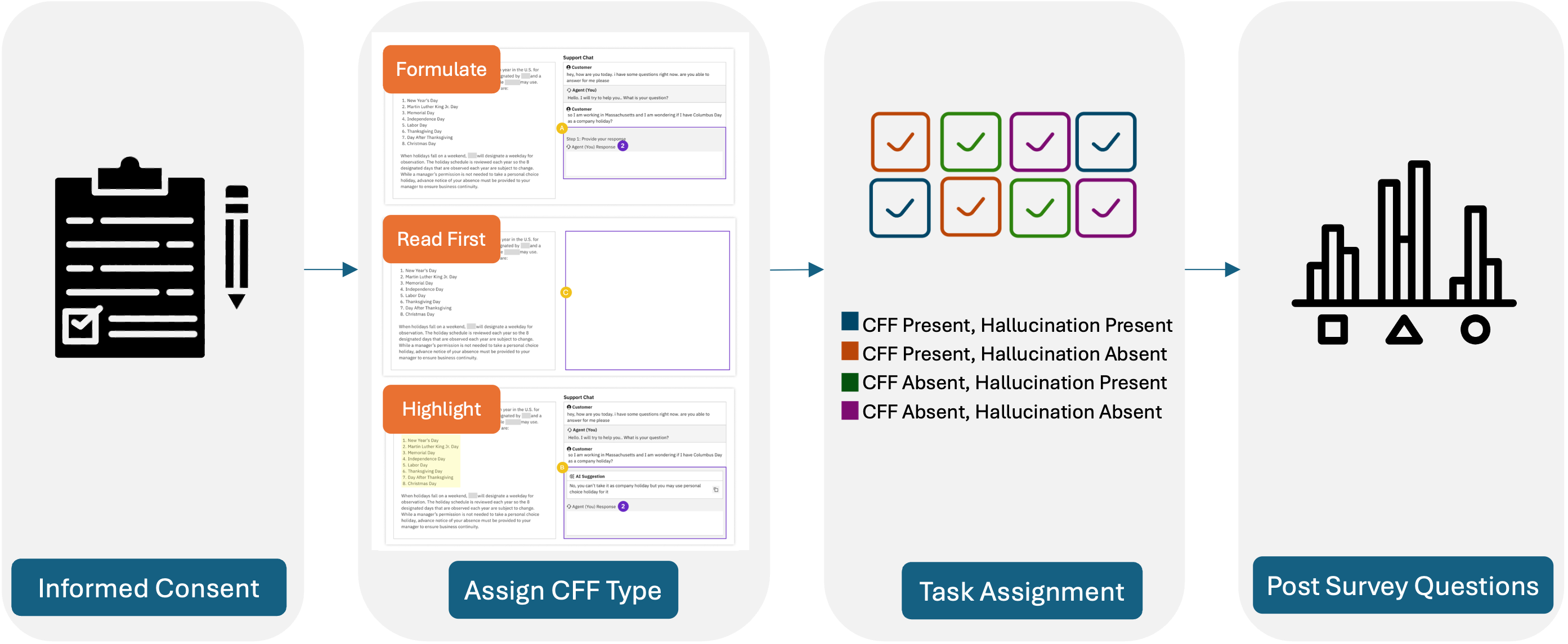}
  \caption{Flow of tasks presented to each user. Each participant was assigned to one of the CFF conditions (Formulate, Highlight, Read First) and then completed 8 tasks with ordering randomized. }
  \label{fig:task5}
\end{figure*}

\subsection{Participants}
We recruited 34 participants via multiple slack channels and word of mouth. Participants were asked to complete a short (30 minute), web-based research experiment on Human-AI Collaboration for HR tasks. The study followed company policies on user data. Participation was voluntary, with the option to withdraw anytime. Data was de-identified, stored securely, and analyzed anonymously. Participants consented to the use of de-identified data for current and future research, ensuring confidentiality.  After the completion of the tasks, they were asked to complete a survey. Survey questions can be divided into four categories: trust and exposure in AI in general, trust in AI-generated responses in study, user confidence in their own ability complete tasks, and subjective perception of the Cognitive Forcing Functions with which they interacted. All questions were asked on a 7-point likert scale.

\begin{figure*}[ht!]
  \centering
\includegraphics[width=0.75\linewidth]{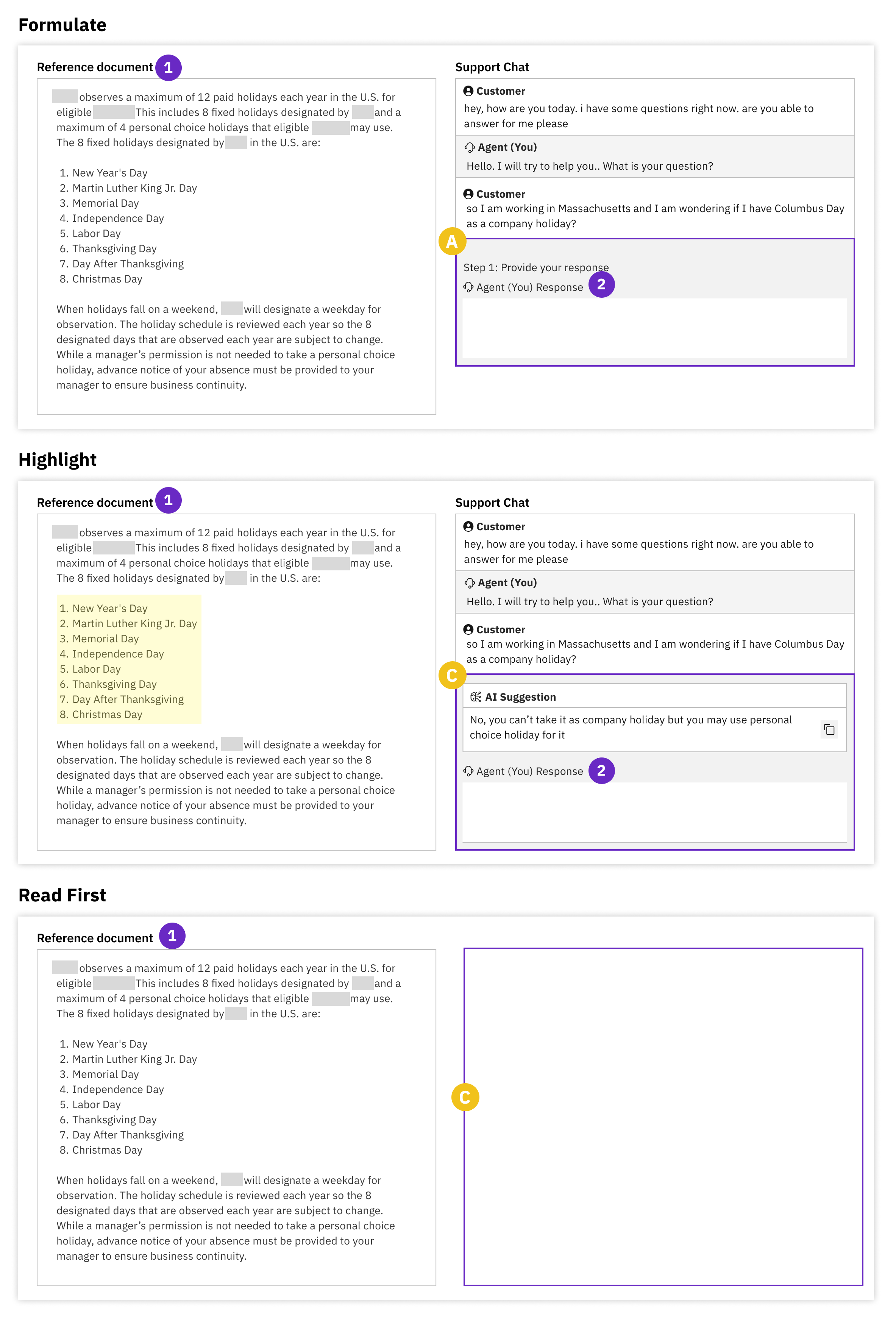}
  \caption{Cognitive Forcing Functions. 1) Reference Document, 2) Textbox for user to submit final response, A) Represents the interaction in the Formulate conditions in which a user is asked to first respond to the question without an AI suggestions. B) Response box corresponding to Highlight Condition, C) Read First condition: User first sees reference document before seeing respective chat and customer query. }
  \label{fig:cff}
\end{figure*}

\subsection{Procedure}
Participants were first presented with a brief description of the study and provided informed consent. Participants were given instructions on the task, as seen in Figure \ref{fig:task}. The study employed a mixed between-within design, with one between-subject variables: CFF Type (Formulate, Read-First, Highlight), and two within-subject variables was whether the CFF was made visible to participants during the task and whether the task contained a hallucination. All participants were presented with the same tasks in a random order to account for potential ordering effects and an equal number of hallucinations as seen in Figure \ref{fig:task5}. Participants were randomly assigned to one of the four Cognitive Forcing Function conditions: Formulate, Read-First, or Highlight. The Cognitive Forcing Functions were designed to provide different types of support to participants in improving the quality of the AI-generated responses. Images of how these different CFFs were implemented can be seen in Figure \ref{fig:cff}. An example image of how the task appeared without the CFF can be seen in Figure \ref{fig:task}. Below, we provide a description of these Cognitive Forcing Functions to improve the quality of the AI-generated responses:
\begin{figure}[ht]
  \centering
\includegraphics[width=1\linewidth]{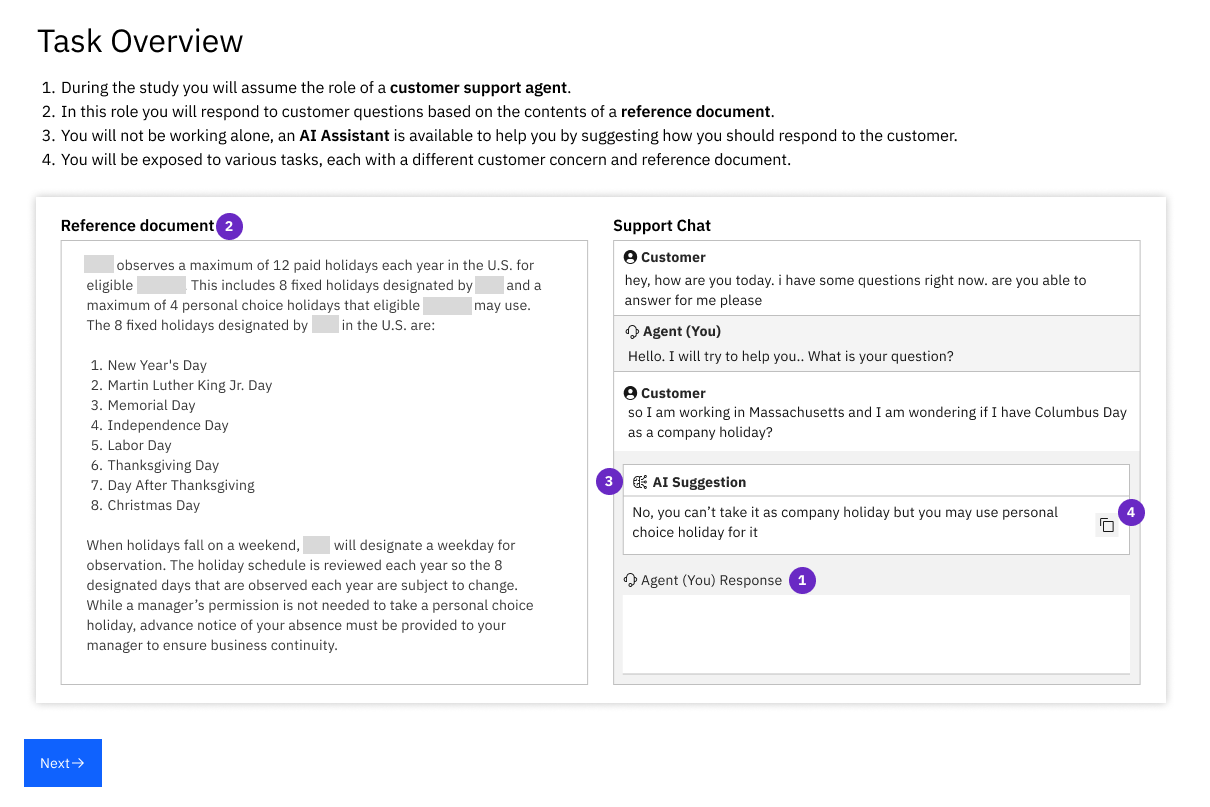}
  \caption{Task overview presented to participants before completing the eight tasks. 1) Refers to the final response submitted by user, 2) Reference document, 3) AI-Generated Suggestion. 4) Refers to copy button presented to users. }
  \label{fig:task}
\end{figure}

\begin{enumerate}
\item \textbf{Formulate} This function required participants to independently construct a response to query without AI assistance before seeing the AI's response. This was intended to encourage participants to think critically about the prompt and not rely solely on the AI's output. Once they submitted their own text, they were presented with both their own response and the AI-generated text. They were able to make their final submission either keeping their own response, using the AI-generated text or editing to submit a combination of both/neither. 


\item \textbf{Highlight} This function provided attention support to the participant by showing the semantic similarity between the model response and coherent sentences within the reference document as a colored overlay. Similarity was calculated via embeddings generated by a sentence transformer. This tool was intended to help participants quickly gauge the quality and faithfulness of the AI-generated response relative to the reference document.

\item \textbf{Read First} This function first showed users the document in which the conversation was grounded and asked the user to read the conversation before seeing the entire task. Once the user has reviewed the reference document they proceed to view the entire task (similar to C in Figure \ref{fig:cff}). 
\end{enumerate}

These Cognitive Forcing Functions were designed to provide different types of support to participants in improving the quality of the AI-generated responses, and were randomly assigned to participants in the study. \za{Participants were instructed to:} \textit{Make sure that your response to the customer is faithful to both the reference document and the prior conversation, and contains correct information that addresses their question or concern. Note that the AI may not always be accurate, so verify against the reference document. }

Participants completed 8 tasks, each with a different question and AI agent response. Half of the tasks included hallucinations randomly inserted throughout the interaction (see Figure \ref{fig:task}) to simulate errors in the AI-generated response and half included the presence of Cognitive Forcing Functions. In each interaction, participants were given the option to edit the AI agent's response to the question to improve it. The order of tasks within each block was randomized to control for order effects. Participants completed a brief post-study survey to collect their subjective evaluation of the Cognitive Forcing Functions. We also logged various user interaction data for analysis. This data included the duration participants spent  completing a task. We also logged copy events, which denote the frequency with which users selected the copy button located next to the AI-generated text during each task (See 4 in Figure \ref{fig:task}).


\subsection{Experimental Setup and Artifacts}
The tasks in our experiment were selected from HR support documents used at the company where the authors are employed. Each task has a related reference document (grounding document), and a simulated conversation between a customer and an agent focusing on the content of the document. The conversations were previously created by five annotators ranging from 6-10 turns and have been segmented at a point where the customer had asked a question. One annotator played the role of a customer and one annotator played the role of the support agent. The final turn of the conversation includes AI-generated responses from Flan-T5-XXL \cite{https://doi.org/10.48550/arxiv.2210.11416}, along with a fine-tuned version optimized for content-grounded dialogue.

Two of the authors reviewed 302 conversations, their corresponding AI-generated responses, their corresponding HR reference documents and selected 8 final tasks spanning topics on maternity leave, health insurance benefits after being rehired by another company after retirement, how to send a provisional  reimbursement form, vacation day policies, air travel policy, booking hotels for personal-use travel policy, personal leave of absence, and health care benefits. Four of the selected tasks had hallucinated responses generated by the model (Retire-Rehire, Birth, Hotel FAQs and Personal Leave of Absence from Table \ref{tab:mytable}) and four the selected tasks had complete and accurate responses generated by the model (Provisional Reimbursement Form, Mental Health Care, Air Travel Policy, and Vacation). \za{The tasks and resulting last turn were selected to include four hallucinated responses and four non-hallucinated responses for our controlled experiment.}

\label{sec:dataQ}
\subsubsection{\za{Reference Document}}
\za{In this study, eight unique reference documents were used. These documents, which are actual materials used by employees of a large multinational company, were designed to assist employees on varying topics including travel rules, leave of absence policies, and maternity benefits. On average, they contained 555.25 words (M = 555.25, SD = 274.41), and each document was structured into 3-5 subsections.  For example, the ``Birth'' reference document is meant to provide guidance to new parents in navigating various administrative and benefits-related procedures following the birth or adoption of a child. It provides information on healthcare and insurance procedures, compliance with benefits policy, financial and insurance considerations, support services, and additional administrative actions.}

\subsubsection{\za{Gold Responses}}
Designing ground truth is a challenging iterative task \cite{muller2021designing}. To generate gold responses for each task, three of the authors independently reviewed the conversations and respective reference documents and generated a gold response. \za{The gold responses generated independently must correctly address the question, be complete, and only include relevant information.} The annotators (three authors) met to compare and contrast their gold responses and resolved any disagreements. The resulting gold responses \za{for all eight tasks} only included relevant and complete information the reference document that addressed the customer's query. The relevant turns of the conversation and resulting gold responses \za{for a subset of the tasks} can be seen in Table \ref{tab:mytable}.

\subsubsection{Rubric Creation}
The authors had extensive discussions with a subject matter expert directly involved in data collection efforts for fine-tuning a model in the HR context to determine how to evaluate the resulting responses from users.  Based on these discussions, we created 6 yes/no questions to determine data quality that can be categorized into three different areas: Faithfulness, Accuracy, and Completeness. \za{These areas were influenced and adapted from Grice's maxims which stress clarity (manner), relevance (relation), truthfulness (quality), and informativeness (quantity) \cite{dale1995computational,jacquet2019impact} and were reduced to yes/no questions to prevent survey fatigue and improve consistency of evaluations.} We also asked an additional yes/no question about use of the AI-generated text, a response later used to determine reliance on the AI-generated text in each task. This question aimed to determine whether participants incorporated any part of the AI-generated suggestion into their final response. In some cases, the final response was an exact copy of the AI suggestion or included only minor edits. In other cases, the influence was more subtle, with specific phrases or ideas from the AI suggestion appearing in the final response. This was measured as a binary metric, indicating whether any portion of the AI-generated text was present in the user's final response. The questions are listed in Table \ref{tab:human}

\begin{table*}[ht]
    \centering
    \caption{Evaluation Criteria For Data Quality and Reliance}
    \label{tab:human}
    \small
    \begin{tabular}{p{1.6cm}|p{9.5cm}|p{1.5cm}}
        \hline
        \textbf{Facet} & \textbf{Question} & \textbf{Metric} \\
        \hline
        Faithfulness & The user response contains information from the reference document & Data Quality  \\
            & The user response contains information outside of the reference document & Data Quality \\
            \hline
        Accuracy & The user response is accurate (when compared to the Gold Response) & Data Quality \\
            &The user response addresses the question in the customer-agent interaction. & Data Quality \\
                   \hline
        Completeness &The user response contains all of the content in the gold response. & Data Quality \\
            &The user response is missing key information included in the gold response. & Data Quality  \\
                   \hline


        Use of AI & The user appears to have used the AI suggested response to formulate their response. & Reliance \\
        \hline
    \end{tabular}
\end{table*}


To determine the quality of the data generated by users in the tasks, the tasks were divided into four annotation units that were then labeled by four of the authors who rated the resulting human-AI responses as ``yes'', ``no'', or ``I don't know'' based faithfulness, accuracy, and completeness. Our expertise in this subject stems from our extensive discussions on task selection, which involved reviewing over 300 HR documents and their respective conversations, as well as co-creating gold standard responses for each task. Prior work has recruited subject matter experts for labeling data to yield more reliable data and has demonstrated that in some contexts, SME annotated data is more reliable than crowsdsourced data \cite{shing2018expert,snow2008cheap}. We spent a considerable time reviewing HR documents for selection in this study and thus had  domain knowledge. By labeling the resulting responses, we were able to have discussions around disagreements and refinement of definitions of categories in our rubric.  Items that were marked as ``I don't know'' were revisited by the team and decisions made based on all of the labeler's agreement. Upon completion of the labeling, another author inspected all evaluations for consistency. In each rubric, evaluators saw the reference document, the gold standard response, the AI generated response, and the submitted response by the user. When evaluating, the conditions for each item to be evaluated were hidden. Evaluators were asked to compare the user response to the gold response. 

\za{The items were developed based on extensive discussions with subject matter experts which led to the decision to use fewer items. Our choice of a three-way rating system ("yes", "no", "I don’t know") both avoids forcing raters into a binary decision when uncertain and reduces complexity of likert scale \cite{company2018go}. Our approach acknowledges the nuanced nature of AI-generated text evaluation. The constructs being measured are specific and narrowly defined, allowing us to capture the essence of what is being measured. We were also cognizant of survey length which can impact quality of responses. Since we surveyed multiple constructs, limiting the number of items per construct kept the survey concise to maintain quality.  Furthermore, the annotators had extensive experience with HR documents and made sure to discuss ratings with the team when any uncertainties arose.}

\section{Results}
\subsection{Participant Demographics and Self Reported Measures}
We recruited 34 participants and asked them details about their roles at the company, and their relationship with AI as well as their mental load with respect to the features with which they were presented. 23.5\% of the participants reported they worked for Human Resources, 17.6\% reported they were a technical lead/manager, 11.7\% reported they worked as a Software Engineer, 5.8\% reported they worked as a researcher. Other roles of participants (> 3\%) included: data scientist, consultant, designer, hardware development, procurement, data governance analyst, client manager, program manager, technician, solutions architect, systems and business analyst, and data labeler. Of the participants 53\% had a Bachelor's Degree, 32\% had a Master's Degree, 7\% had a Ph.D. or higher and 8\% reported having another degree which consisted of: nursing degree, jurisdoctorate, and highschool diploma.

\begin{table*}[t]
\small
    \centering
    \caption{Examples of gold responses for a subset of tasks.}
    \begin{tabular}{p{3cm}p{4cm}p{6cm}}  
        \toprule  
        \textbf{Task Name} & \textbf{Relevant Turns of Conversation} & \textbf{Gold Response} \\
        \midrule  
Vacation&\textbf{Customer}: Yay, that is good news. If I change to a flexible work schedule after I am married, will that affect my vacation eligibility?&
Yes that will affect your vacation eligibility. Employees on flexible work week schedules earn vacation based on hours worked per week and years of service.\\
Birth&\textbf{Customer}: How many days of maternity leave do I get?&
I do not know how many days you get off for maternity leave. You must speak to your manager.\\
Personal Leave of Absence&\textbf{Customer:} this is another question I want to know so that i can take off partial of the day with the justification&
It sounds like you’re asking about taking half a day off which is not a leave of absence. Authorized time off for 10 days or less is considered time off, but it is not a leave of absence.\\
        \bottomrule  
    \end{tabular}
    \label{tab:mytable}
\end{table*}
\subsubsection{User Trust and Exposure to AI}
Participants in the study rated their exposure to Artificial Intelligence on a Likert scale from 1 (``I have never heard of AI'') to 5 (``I have extensive experience in AI research and/or development.''), with an average exposure level of 3.39 (SD = 0.92), indicating a moderate level of familiarity. Most respondents fell within the moderate exposure range (Median = 4). Participants responded to ``In general, to what extent do you believe that AI systems are capable of generating accurate text in conversational contexts?'' on a 1 to 7 scale  with (M=5.06, SD=1.18) with most participants skewing to a high level of familiarity (median=5). Participants also responded to ``In general, how likely are you to trust text generated by AI systems?'' on a 1 to 7 scale with (M= 4.41, SD=1.3).

\subsubsection{Trust in AI-generated Responses in Study}
Participants in the study were asked to rate the AI-generated responses in the tasks they completed on a scale of 1 (Never) to 7 (Always) for accuracy (M=4.36, SD=1.2), completion (M=4.20, SD= 1.3), and usefulness (M=4.68, SD=1.36). Participants skewed to rating the AI generated responses highly for accuracy (median=4), completeness (median=4), and usefulness (median=5). 

\subsubsection{User Self confidence in Ability to Complete Tasks}
Participants in the study were asked to rate their confidence in their own ability to to provide answers in the tasks on a scale of 1 (Never) to 7 (Always) for accuracy (M=5.84, SD=1.12). Participants skewed to rating themselves highly for their perceived ability to produce accurate (median=6) and conversational (median=6) responses.

\subsubsection{Subjective Perception of Cognitive Forcing Functions}
Participants in the study described why they felt the feature was helpful/unhelpful. For coding these responses, two authors independently reviewed the responses to generate the codebook for each Cognitive Forcing Function. The coders then met to resolve any differences. Similar codes were consolidated, and codes identified by only one coder were discussed until a consensus was reached.  The results can be seen in Table \ref{tab:dislikes}.

\begin{table*}[ht]
    \centering
    \caption{Why Users Liked/Disliked Cognitive Forcing Function}
    \small
    \begin{tabular}{p{1.5 cm}|p{1.5cm}|p{4.5cm}|p{5.5cm}}  
        \toprule  
\textbf{Preference}& \textbf{CFF}&\textbf{Reason for Preference}&\textbf{Example} \\
        \midrule
        \textbf{Like} & \textbf{Read First}  &Perceived Time Pressure & \textit{It gave me time to pause and read. I know there wasn't a real person behind the chat nor was there any time constraints, but it made it feel more urgent to have to read and think about responding at the same time.} \\\cline{2-4}
   &\textbf{Highlight}& Relevant Passages Outlined &\textit{The feature was helpful by outlining the relevant passage}\\
      &&Made User More Efficient &\textit{Able to quickly read the relevant text without reading the entire article} \\  
       &\textbf{Formulate}  &Helpful for Confirmation &  \textit{The feature was helpful for me to confirm my results but the answers were not always correct.} \\ 

 \hline
      \textbf{Dislike} &\textbf{Read First}&  Repetitive & \textit{it was repeated in the next page}\\ \cline{2-4}
&\textbf{Highlight} &Irrelevant Passage Highlighted& \textit{The highlighting in nearly all cases did not select the part of the document which contained the answer.}\\ \cline{2-4}
&\textbf{Formulate}& More Work for User &\textit{It's always easier to edit someone's work than to start from scratch... I was more quickly able to review the suggested response, check the reference doc for accuracy and give an accurate response than if I had to answer on my own}\\

        \bottomrule 
    \end{tabular}
    \label{tab:dislikes}
    \end{table*}

\subsection{\za{Data Quality in Human-AI Collaborative Content Grounded Data Generation \textbf{(RQ1)}}}

To address our research questions, we conducted a linear mixed-effects model (LMM) analysis to examine the relationships within the dataset while accounting for repeated measures at the task level.
Mixed-effects models are particularly useful for handling hierarchical data structures \cite{schielzeth2013nested} and incorporating random effects \cite{bolker2015linear}, making them well-suited for our study, where each participant completed multiple tasks. \za{Our approach focused on assessing the impact of Cognitive Forcing Functions (CFF) (between-subjects), hallucination presence, and CFF presence (within-subjects) on dependent variables such as data quality and use of AI. To appropriately model repeated observations per user, we included random intercepts for each participant, allowing us to account for individual variability in baseline performance while preserving the structure of the 8 tasks per user. We used ANOVA-like significance testing on the fixed effects in our LMM by applying F-tests with Satterthwaite’s approximations for degrees of freedom, which provide more accurate significance estimates in mixed models by accounting for  hierarchical dependencies.}

\begin{figure*}[ht]
  \centering
\includegraphics[width =1\linewidth]{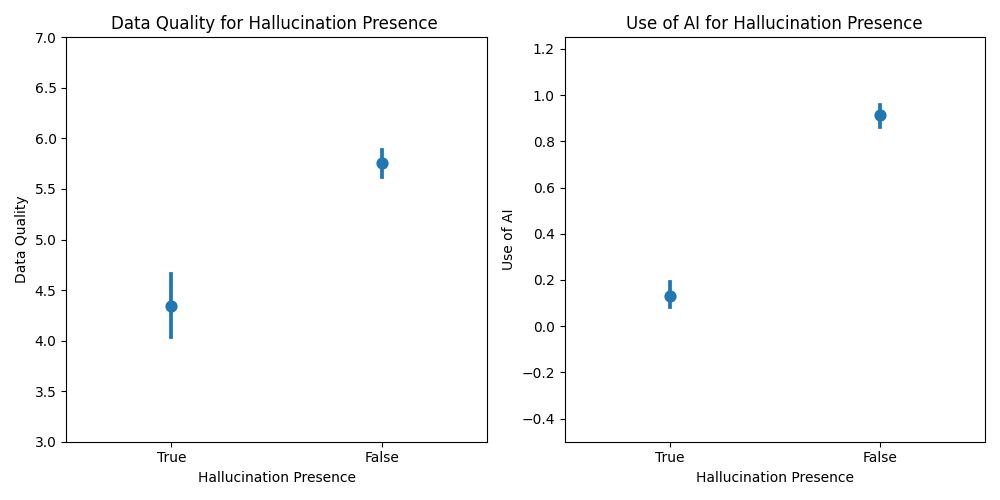}
  \caption{Impact of Hallucinations on Data Quality and AI Use. Data quality decreases (left) and AI reliance drops (right) when hallucinations are present. }
  \label{fig:hallucination}
\end{figure*}

In our design, CFF Type refers to the type of Cognitive Forcing Function (CFF) assigned to a task (Formulate, Read First, or Highlight). However, this assignment does not necessarily indicate that the intervention was applied. We also define a separate variable, CFF Presence, which is a binary indicator (True or False) of whether the assigned CFF was actually presented to the participant during the task.  Hallucination indicates whether a hallucination is present in the AI-generated text presented in a task (Present, Absent) and CFF Presence indicates whether the Cognitive Forcing Function is present in a task (True, False).  The dataset consisted of 272 observations, with each observation belonging to a unique group represented by the combination of user and specific task. Table \ref{tab:anova-results} shows the results of our mixed linear model regression of Hallucinations and Cognitive Forcing Functions on data quality to address \textbf{RQ1}. We find that hallucinations have a significant negative effect on data quality (\textit{F}(1, 229) = 73.07, \textit{p} < .001), indicating that the presence of hallucinations substantially reduces data quality. Tukey Post-hoc analysis confirms this effect, showing that when hallucinations were present, data quality was significantly lower (\textit{M} = 4.34, SE = 0.137) compared to when hallucinations were absent (\textit{M} = 5.77, SE = 0.137), with a mean difference of -1.43 (\textit{t}(229) = -8.548, \textit{p} < .0001) (see Figure \ref{fig:hallucination}). Therefore, we accept H1.  

Neither Cognitive Forcing Functions (CFF) (\textit{F}(2, 31) = 0.19, \textit{p} = .83) nor CFF presence (\textit{F}(1, 229) = 3.12, \textit{p} = .079) had a statistically significant main effect on data quality. Additionally, no significant interaction effects were observed between CFF, CFF presence and hallucinations (\textit{F}(2, 229) = 1.870, \textit{p} = .15), suggesting that the presence of a CFF did not mitigate the negative impact of hallucinations on data quality. Thus, we reject H2A-H2C since we do not see any statistically significant interactions between CFF Type x Hallucination x CFF Present. 




\begin{table*}[ht]
    \centering
    \caption{Analysis of Variance Table for Cognitive Forcing Functions (CFF), Hallucinations, and CFF Presence on Data Quality}
    \label{tab:anova-results}
    \begin{tabular}{lcccc}
        \toprule
        \textbf{Effect} & \textbf{Num DF} & \textbf{Den DF} & \textbf{F Value} & \textbf{p-value} \\
        \midrule
        CFF & 2 & 31 & 0.187 & 0.830 \\
        Hallucination & 1 & 229 & 73.067 & \textbf{<0.001} *** \\
        CFF Presence & 1 & 229 & 3.116 & 0.079 . \\
        CFF x Hallucination & 2 & 229 & 0.652 & 0.522 \\
        CFF x CFF Presence & 2 & 229 & 1.254 & 0.287 \\
        Hallucination x CFF Presence & 1 & 229 & 0.441 & 0.508 \\
        CFF x Hallucination x CFF Presence & 2 & 229 & 1.870 & 0.157 \\
        \bottomrule
    \end{tabular}
    \begin{center}\textit{Significance codes:} 0 ‘***’ 0.001 ‘**’ 0.01 ‘*’ 0.05 ‘.’ 0.1 ‘ ’ 1. \end{center}
\end{table*}

\begin{table*}[ht]
    \centering
    \caption{Analysis of Variance Table for Cognitive Forcing Functions (CFF), Hallucinations, and CFF Presence on Use of AI Response.}
    \label{tab:anova-use-ai}
    \begin{tabular}{lcccc}
        \toprule
        \textbf{Effect} & \textbf{Num DF} & \textbf{Den DF} & \textbf{F Value} & \textbf{p-value} \\
        \midrule
        CFF & 2 & 31 & 0.566 & 0.573 \\
        Hallucination & 1 & 229 & 463.519 & \textbf{< 0.001} *** \\
        CFF Presence & 1 & 229 & 0.046 & 0.831 \\
        CFF x Hallucination & 2 & 229 & 0.366 & 0.694 \\
        CFF x CFF Presence & 2 & 229 & 4.944 & \textbf{0.008} ** \\
        Hallucination x CFF Presence & 1 & 229 & 0.381 & 0.538 \\
        CFF x Hallucination x CFF Presence & 2 & 229 & 2.164 & 0.117 \\
        \bottomrule
    \end{tabular}
     \begin{center} \textit{Significance codes:} 0 ‘***’ 0.001 ‘**’ 0.01 ‘*’ 0.05 ‘.’ 0.1 ‘ ’ 1. \end{center}
\end{table*}

\subsection{\za{Reliance in Human-AI Collaborative Content Grounded Data Generation \textbf{(RQ2)}}}

Table \ref{tab:anova-use-ai} shows the results of our mixed linear model regression of Hallucinations and Cognitive Forcing Functions on AI usage. The main effect of Hallucination is significant, causing a decrease in the use of AI suggestions (\textit{F}(1, 229) = 463.519, \textit{p} < .001). Tukey post-hoc analysis confirms this effect, showing that when hallucinations were present, AI usage was significantly lower (\textit{M} = 0.133, SE = 0.0299) compared to when hallucinations were absent (\textit{M} = 0.916, SE = 0.0299). The difference between conditions was substantial (\textit{t}(229) = -21.529, \textit{p} < .0001). This finding suggests that users did not rely or use the AI-generated responses when they consisted of hallucinations, so we accept H3. 

\begin{figure}[t]
 \centering
\includegraphics[width=1\linewidth]{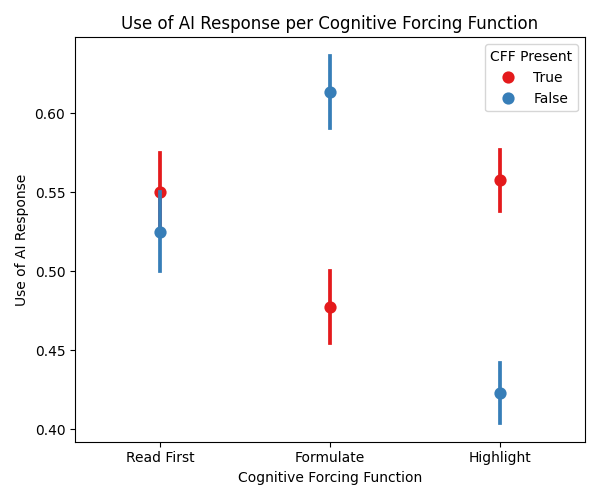}
  \caption{Use of AI by Cognitive Forcing Function Type. When the Cognitive Forcing Function was absent, participants in the Formulate condition used AI significantly more than those in the Highlight condition.}
  \label{fig:useCFF}
\end{figure}

Although CFF Presence had a significant interaction with CFF (\textit{F}(2, 229) = 4.944, \textit{p} = 0.008), this effect does not provide evidence supporting H4A-H4C, which predicted that CFF and CFF Presence would mitigate the negative impact of hallucinations on AI reliance. The three-way interaction between CFF, Hallucination, and CFF Presence (\textit{F}(2, 229) = 2.164, \textit{p} = 0.117) failed to reach statistical significance, indicating that CFF and CFF Presence did not significantly influence how hallucinations affected AI reliance. Therefore, we reject H4A-H4C, as our findings indicate that the presence of a CFF does not significantly alter how users respond to hallucinations in AI-generated outputs. However, Tukey post-hoc comparisons for the significant CFF Presence × CFF interaction reveal that when Cognitive Forcing Functions were absent, participants used AI significantly more in the Formulate condition (\textit{M} = 0.614, SE = 0.0523) compared to the Highlight condition (\textit{M} = 0.423, SE = 0.0481). The difference between these conditions approached significance (\textit{t}(74.5) = 2.684, \textit{p} = 0.0907), suggesting a trend where participants relied more on AI suggestions in the Formulate condition when no CFF was present (see Figure \ref{fig:useCFF}).

\subsection{\za{What types of reliance behaviors are observed in Human-AI Collaborative Content Grounded Data Generation? (RQ3)}}
\label{sec:r4qb}
We define reliance on AI-generated text, including overreliance and underreliance, drawing from prior work, which defines reliance as accuracy on correct or incorrect AI advice \cite{10.1145/3411764.3445717}, ratio of reliance on incorrect AI advice \cite{buccinca2021trust}, ratio of reliance on correct or incorrect AI advice \cite{jakubik2022empirical}. We take a similar approach. A user's reliance on AI-generated text can be classified as overreliance or underreliance based on their interaction with hallucinated and non-hallucinated AI responses. Overreliance occurs when a user incorporates hallucinated AI-generated text, meaning they rely on incorrect AI advice. Underreliance occurs when a user disregards factual and faithful AI-generated text, failing to leverage correct AI advice. To categorize users as overreliers or underreliers, we examined their behavior across all eight tasks they completed, four of which contained hallucinated AI responses and four that did not.  

To define overreliance, we calculated the median number of times users incorporated AI-generated text when the response was hallucinated (median = 0, range = 0 to 4). Users who used hallucinated AI responses more frequently than the median were classified as overreliers. Similarly, to define underreliance, we calculated the median number of times users ignored AI-generated text when the response was non-hallucinated (median = 0, range = 0 to 4). Users who never used non-hallucinated AI responses were classified as underreliers. Since reliance was assessed separately for hallucinated and non-hallucinated responses, it was possible for a user to be classified as both an overrelier and an underrelier. Appropriate reliers were those who neither overrelied nor underrelied on AI-generated text. These definitions align with prior work on reliance behavior \cite{buccinca2021trust,10.1145/3411764.3445717,jakubik2022empirical}. Across our participants, 13 were classified as underreliers, 9 as appropriate reliers, 6 as both overreliers and underreliers, and 6 solely as overreliers.

\begin{figure}[t]
\centering
\includegraphics[width=1\linewidth]{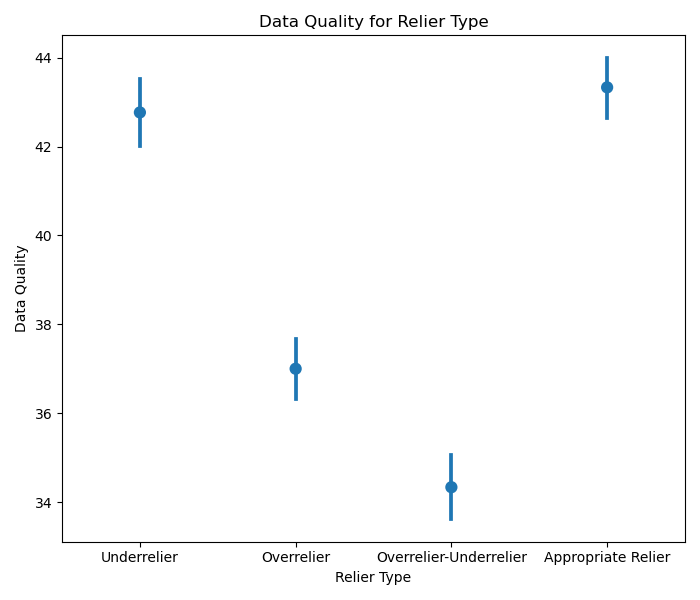}
  \caption{Data Quality by Reliance Type: Data quality scores for under-reliers, over-reliers, combined over/under-reliers, and appropriate reliers.}
    \Description{Data Quality by Reliance Type: Data quality scores for under-reliers, over-reliers, combined over/under-reliers, and appropriate reliers.}
  \label{fig:reliance}
\end{figure}

\begin{itemize}
\item \textbf{Overrelier}: Used an AI response that is hallucinated one or more times. Used non-hallucinated AI response 1 or more times.  
\item \textbf{Underrelier}: Never used non-hallucinated AI response. Never used hallucinated response. 
\item \textbf{Overrelier-Underrelier} Used an AI response that is hallucinated one or more times. Never used non-hallucinated AI response. 
\item \textbf{Appropriate Relier}: Used non-hallucinated AI response at least once. Never used non-hallucinated response.  
\end{itemize}
We conducted an ANOVA to examine the effects of reliance type, CFF, and their interaction on the sum of data quality for all tasks completed by a user.  The results revealed a statistically significant effect of reliance type on data quality F(3,268)=110.05, p<0.0001. This suggests that different levels of reliance significantly influence data quality scores. Further Tukey post-hoc analysis revealed several significant pairwise comparisons: Appropriate reliers (M=43.33) had higher quality scores than overreliers (M = 36) p < 0.01, and overreliers-underreliers (M=34.33) p< 0.01.  Underreliers (M=42.77) also had higher quality scores than both overreliers and overreliers-underreliers p< 0.05.

\subsubsection{Patterns of User Overreliance on AI-Generated Text}
The gain a better understanding of how exactly users were \za{relying on the} AI-generated text \za{(RQ3)}, we analyzed the data and coded the distinct ways we observed users were relying on the AI response when hallucinations were present and when hallucinations were absent. For coding, two authors independently reviewed the data to generate the codebook and then met to resolve any differences. Similar codes were consolidated, and codes identified by only one coder were discussed until a consensus was reached. This resulted in five ways users were relying on the AI response when the AI was hallucinating: exact copy, copy and minor edit, exact copy and append with correct response, copy and removal of information, copy and added content from the reference document. We logged copy events, but found that using the copy button was not always a reliable indicator of AI-generated text usage, as demonstrated in the example below. In this case, the user never selected the copy button, yet they heavily relied on the AI-suggested response to compose their final answer. Various other strategies for incorporating AI-generated text, along with additional examples, are detailed in Table \ref{tab:copies}.
 
\begin{description}
\label{sec:formulate}
\item\textbf{User’s response before exposure to AI-generated suggested response:} I am sorry but I do not have that information. (User 220, Formulate)
\item\textbf{User’s final response after exposure to AI-generated suggested response:} You must sign up for benefits with your new employer since \ul{your benefits may be reduced or eliminated depending on your income.} (User 220, Formulate)
\item\textbf{AI-generated suggested response:} Yes, you have to. If not, \ul{your benefits will be reduced or eliminated depending on the amount of income you earn.} 
\end{description}

\begin{table*}[ht]
    \centering
    \caption{Overreliers: Strategies employed when using AI response when Hallucination Present for Task: Hotel FAQs}
    \begin{tabular}{p{7cm}p{7cm}}  
        \toprule  
\textbf{Category}&\textbf{Example} \\
        \midrule
     
      Exact Copy & \textbf{Yes of course you may use [Corporate Credit Card]!}\\
     
        Copy and Minor Edit & Yes of course you may use [Corporate Credit Card] card that was provided to you \\
        
        Copy and Removal of Information& Yes you may use [Corporate Credit Card]\\ 
        Exact Copy and Append with Correct Response &  Yes of course you may use [Corporate Credit Card]!  Hotel reservations for personal travel may be made directly with the hotel property or with the [redacted] designated travel reservation system using your personal credit card \\ 
         Copy and Added Content from Reference Document (Not Correct Response) & Yes of course you may use [Corporate Credit Card]!  Please make sure to review all travel polices and FAQs at [redacted]\\
        \bottomrule 
    \end{tabular}
    \label{tab:copies}
    \end{table*}

\subsubsection{Patterns of User Underreliance on AI-Generated Text} 
We also examined the strategies employed by participants who were provided with a factual and faithful AI-generated response but chose not to use it. Analyzing cases where users opted against incorporating non-hallucinated AI text, we found that many instead copied text directly from the reference document, regardless of its relevance to the user query. This strategy often led to unnecessarily long responses that failed to adequately address the customer's question. A common pattern in these responses was the inclusion of extraneous information, making them both less concise and less effective.

Additionally, we found that simply copying non-hallucinated AI-generated text does not always result in high-quality responses. While appropriate reliers produced the highest-quality responses on average (M = 43.33) compared to overreliers (M = 36) and underreliers (M = 42.77), there were instances where reliance on factual AI text still led to suboptimal outputs. In particular, some users who incorporated AI-generated text also appended excessive or irrelevant content from the reference document, ultimately reducing response quality. An example of this is illustrated below, where the AI-generated text is bolded, and additional superfluous content has been incorporated from the reference document.
\begin{quote}
[redacted]. When Medicare is primary over the [redacted] medical plans for you or for your eligible dependent, [redacted]. You and your eligible dependents are eligible for SHAP benefits whether you are enrolled in an [redacted] retiree medical plan, [redacted] Self-Managed Plan (Hawaii), an HMO, or if you have opted out of the [redacted] medical plans.   [redacted]. Additionally, please note: \textbf{You must complete the SHAP Reimbursement form and submit it on a quarterly basis after the quarter is completed or at the end of the year, but no later than December 31 of the next year}
\end{quote}

\begin{table*}[ht]
    \centering
    \caption{Human Judgement "Used AI Suggestion" and System Copy-Event  Overlap with Examples from Retire-Rehire Task.}
    \label{tab:overlap}
    \small
    \begin{tabular}{|p{1.5cm}|p{1.5cm}|p{2.5cm}|p{6cm}|}
            \toprule  
\textbf{Used AI }&\textbf{Copied} &\textbf{Number of Tasks} &\textbf{Example}  \\
\toprule  
Yes& Yes & 97& Yes, you have to. If not, your benefits will be reduced or eliminated depending on the amount of income you earn\\
\hline
Yes& No & 45 & Yes, your new employer's plan will become your primary medical coverage \\
\hline
No& Yes & 10& No, not quite necessary, if your spouse has a health coverage for you\\
\hline
No& No & 120& Your retirement benefits will continue with no change\\
\bottomrule
    \end{tabular}
\end{table*}





\section{Discussion}

\subsection{Unintended Consequences of Cognitive Forcing Functions on AI Use}
The negative impact of hallucinations on data quality is well-documented \cite{leiser2023chatgpt} —incorrect AI recommendations have been shown to harm users in decision-making contexts \cite{de2020case}. However, our study revealed that cognitive forcing functions (CFFs) were not successful in mitigating this negative impact, unlike their effectiveness in reducing overreliance in prior studies \cite{buccinca2021trust}. In the condition where users had to compose their own response before seeing the AI-generated response (Formulate), which was adapted from the decision making context in \cite{buccinca2021trust}, we found that users relied on the AI more when they were not presented with the CFF. In \cite{buccinca2021trust}, requiring users to make a decision before viewing the AI’s recommendation successfully mitigated overreliance, but participants disliked the process. 

In our study, which involved more complex tasks than just decision making (e.g., reading a prior conversation, reviewing a context document, and generating a correct response), users who were not presented with a CFF in the Formulate condition—meaning they received the AI-generated response immediately—were more likely to incorporate AI suggestions into their final answers. One possible explanation comes from user feedback: the CFF that required them to generate their own response first was perceived as too effortful. As a result, when this extra effort was removed, participants were more inclined to rely on AI-generated content, possibly as a way to reduce cognitive load.  This finding serves as a cautionary tale: CFFs that impose additional effort may lead users to compensate by relying more on AI in subsequent tasks. To address this, we recommend implementing CFFs conditionally—specifically in cases where the AI does not display high confidence. This conditional approach can help balance the benefits of CFFs while minimizing unintended shifts in user behavior.

\subsection{New Patterns of AI Reliance in Co-creation}

In classic decision-making contexts, reliance on AI is typically categorized using a 2 × 2 matrix. One axis represents whether the AI agent’s recommendation is correct or incorrect, while the other represents whether the human decision-maker accepts or rejects the AI’s decision \cite{buccinca2021trust,10.1145/3579605}. This framework defines three types of reliance: underreliance, overreliance, and appropriate reliance. For example, if the AI provides an incorrect recommendation and the human rejects it, they are appropriately relying on their own judgment. However, if the AI’s recommendation is correct and the human still rejects it, this is considered underreliance.  While this framework works well for decision-making tasks, it oversimplifies reliance in co-creative text-generation contexts, such as the one examined in this study. Our findings reveal novel forms of overreliance that have not been previously observed in human-AI decision-making studies. Specifically, we found that users may not only accept incorrect AI-generated responses outright but also append them with additional information—sometimes including correct information. This results in a hybrid response that contains both correct and incorrect information, even when they conflict.  

We also found that underreliers and appropriate reliers produced the highest-quality responses, outperforming both overreliers and those who alternated between overreliance and underreliance. This suggests that, in a co-creative context, underreliance may not be as detrimental as it is in decision-making tasks. However, underreliance in text generation is more complex than simply rejecting correct AI-generated content. Some users compensated by relying heavily on the reference document, copying large chunks of text instead of using AI-generated content. While this approach helped maintain accuracy, it often resulted in lengthy, redundant responses that included irrelevant information and did not always directly address the customer’s query.  

These new forms of AI reliance create challenges for evaluating text, as they go beyond a simple binary decision framework like the 2×2 matrix. Assessing whether a person incorporated AI-generated text into their final response requires considering multiple factors. One straightforward approach is to log explicit copying behaviors. However, this method is limited because AI exposure alone can influence responses, even if the text is not copied verbatim.  

In our study, we observed cases where participants did not use the copy button—an action we logged—but still incorporated portions of AI-generated text into their final responses. To address this, we implemented a manual evaluation process, developing a rubric and engaging in deliberation to assess whether participants relied on AI-generated content. However, manual evaluation is neither scalable nor efficient, as it requires extensive discussion to establish and apply consistent criteria.  A promising future direction is to explore LLM-as-a-judge methods with human oversight for verification. These approaches, particularly as they improve, could help evaluate text for faithfulness and groundedness more efficiently \cite{zheng2023judging}.

\section{Limitations and Future Work} 
While our study offers valuable insights, we acknowledge several limitations. Firstly,  the study's specific focus on conversational customer support might limit its applicability to other domains or use cases of LLMs, as the impact of hallucinations could vary in different contexts. However, we do believe that our findings may generalize to other content-grounded contexts in which a AI-generated text is required to be factual and faithful.  Additionally, our approach to measuring data quality included the creation of gold response which may not be scalable. \za{Our participant pool was limited to internal members of the company employing the authors, due to the confidentiality of the HR documents. Recruitment was conducted through HR-related Slack channels, targeting individuals involved in HR discussions. This approach ensured participants were directly invested and familiar with the material, enhancing the quality of data compared to using external sources like Mechanical Turk workers or other crowdworking platforms.} However, despite only 25\% of our participants reported being from Human Resources, participants on average reported high confidence in their ability to compose high quality data in this study. The relatively small sample size limits the generalizability of our findings, and future work should examine whether these patterns hold across a larger and more diverse participant pool.

\section{Conclusion}

In this paper, we addressed several research questions pertaining to the impact of hallucinations, and Cognitive Forcing Functions on the quality and user reliance observed in conversational data generated through human-AI interactions. When such data is to be used for fine-tuning LLMs in a context like HR or customer support, making sure this data is as high quality as possible is critical.  We have introduced a comprehensive rubric that evaluates AI-generated conversational data based on criteria such as faithfulness, accuracy, completeness and influence of AI. This rubric serves as a practical tool for assessing and enhancing the quality of such data, promoting more meaningful and informative interactions between users and AI agents. 
Our findings underscore the significant impact of hallucinations on the quality of generated data, showing the importance of managing and mitigating hallucinations in AI-generated content to ensure data integrity and reliability. We observed variations in reliance patterns on different Cognitive Forcing Functions.  Additionally, we provide a taxonomy of different types of reliers within the context of human-AI interactions. Notably, our study has identified that both overreliers-underreliers and overreliers tend to underperform, emphasizing the critical role of achieving a balanced reliance on AI-generated responses for optimal user experiences. This contribution enhances our understanding of how hallucinations, human evaluation of data quality, and reliance influence Human-AI Interaction, providing insights for the design of improved systems in this domain. 
\bibliographystyle{ACM-Reference-Format}
\bibliography{sample}



\end{document}